**Hopping mechanism for superconductivity revealed by Density Functional Theory**

*Jose A. Alarco[1] and Ian D. R. Mackinnon[2]**

Jose A. Alarco

[1]School of Chemistry and Physics, and Centre for Materials Science, Queensland University of Technology, Brisbane, QLD 4001, Australia

Ian D. R. Mackinnon

[2]School of Earth and Atmospheric Science, Queensland University of Technology, Brisbane, QLD 4001, Australia

Email: ian.mackinnon@qut.edu.au

[1]ORCID: 0001-6345-071X; [2]ORCID: 0002-0732-8987

Funding: This research received no external funding

**Electronic band structures of $MgB_2$ show that σ bands along the G–A reciprocal direction (parallel to the *c*-axis) in combination with electron-phonon coupling to the $E_{2g}$ mode effect charge transfer. DFT calculation shows that the cosine-shaped σ band along G–A reveals energy asymmetry matching the superconducting gap. Sigma bands parallel to the G–A direction calculated at intervals along the G–M direction show band structure unlike that at high symmetry nodes. For example, bands in close proximity to $E_F$ do not emulate the degeneracy of bands along G–A. Near $E_F$, bands split and align favourably for electron-hole pairing with the nodal inflection point located at $E_F$. With increased isotropic pressure, changes to σ bands near $E_F$ occur with increased intersections of folded Fermi-surfaces and band crossings. Band intersections that define the fraction of coherent nesting at pressure show equivalent estimates for $T_c$ determined by experiment and by the Kohn phonon anomaly. Tight-binding equations, including corrections to describe the observed asymmetry of a cosine-shaped band, show that a hopping mechanism is associated with cosine band asymmetry, the superconducting gap and Fermi surface nesting. Cosine band asymmetry may be inimical to superconductivity mechanisms in multi-element compounds.**

1. Introduction

Constructive design of new superconductor compounds requires use of electronic band structure calculations (EBSs) combined with understanding of real space orbital interactions. Sigma (σ) bands are ideal modalities for this design strategy because they directly connect atomic positions and orbitals, or using alternative concepts of bond hybridisation, they display orbital overlap between atomic positions to establish bonding.[1, 2] The EBS for $MgB_2$ shows σ bands along the Γ–A reciprocal direction[3] which in real space is parallel to the *c* axis. We have shown that the σ band in $MgB_2$ is cosine shaped when calculated with P6/mmm symmetry or as a folded cosine at the inflexion point with a superlattice symmetry (e.g. $P\bar{6}c2$).[4, 5]

Cosine shaped electronic bands are encountered in standard DFT calculations of superconductors.[4, 6] Cosine bands typically located near, or crossing, the Fermi level and linked to high symmetry nodes, display an energy asymmetry with a strong correlation to the experimentally determined superconductivity transition temperature, $T_c$.[4, 7-9] Extensive investigations of this correlation of energy asymmetry with the superconducting gap as a function of pressure and composition for exemplar compounds[4, 7-10] shows that useful mechanistic insight can be obtained from electronic band structures (EBSs).

To explore mechanisms in detail using *ab initio* DFT without adjustment of key parameters or correction factors, a high k-grid resolution is required.[11] With today's software and computational capacity such resolution with band structures is possible and necessary in order to understand phenomena of meV magnitude. $MgB_2$ is a particular case where the $E_{2g}$ phonon is key to superconducting behaviour and necessitates detailed interrogation to identify the superconducting gap using band structures calculated by DFT.[4] With evolution of computational methods and capacity, and demonstration of cosine shaped bands crossing $E_F$ in $MgB_2$,[4] we turn our attention to specific charge/electron transfer mechanisms that can be revealed by high resolution band structures using *ab initio* DFT.

An and Pickett[3] recognised that $sp^2$ hybrid bonds, evident as σ bands in $MgB_2$, effect σ band to π band charge transfer and that the $E_{2g}$ phonon splits the band uniformly along Γ–A. Others[12-15] confirmed the key role of the $E_{2g}$ phonon in the electron-phonon mechanism for superconductivity in $MgB_2$. Following the analysis of deformation potential in $MgB_2$ by An and Pickett,[3] a later study[16] used DFT calculations to examine the deformation of specific atomic bonds and the charge distribution induced by particular vibration modes within the structure. By applying differential atom displacements to the B–B bond in $MgB_2$, links to changes in σ bands, Fermi surface topologies and the net volume between tubular σ surfaces in reciprocal space show strong correlations to the anomaly shown in calculated phonon dispersions (PDs).[5, 16] The tubular Fermi surfaces of light and heavy effective masses are utilised in discussions of hopping mechanisms below.

Understanding the orbital character of hybridised bonds combined with the electronic properties of a compound is essential for explanations of physical properties,[17] and critical for the design and synthesis of new compounds with targeted properties.[18, 19] Experimental observations of superlattice character in $MgB_2$ using Raman spectroscopy[5] and with synchrotron THz spectroscopy[20] led to an examination of the P6/mmm symmetry typically attributed to $MgB_2$ based on atomic positions.[21] The bonding and antibonding character of bands and orbitals are key features that provide a more detailed, and nuanced, indication of superlattice symmetry that is often overlooked, or indeterminate in dynamic systems.[4, 20] For example, spectroscopic data[20] and *ab initio* DFT calculations[4, 5] show that a superlattice structure with reduced symmetry, such as $P\overline{6}c2$, generates Brillouin zone boundary reflections, Umklapp processes, and substantially enhanced nesting relationships.[4] Vibration frequencies that should not be observed according to the P6/mmm symmetry become detectable in spectra, correspond with superlattice crystallography[5, 20] and cosine folding[4, 20] that displays a clear numerical connection to the superconducting gap. In addition, analysis of EBSs and Fermi surfaces allows for direct identification of Cooper pairing and the superconducting gap, particularly when the k-grid resolution of a DFT calculation is suitably calibrated to structural parameters.[4]

Band structure calculations on condensed matter systems with electrons strongly bound to each atom are commonly ascribed to a Tight Binding (TB) model.[6, 20, 22, 23] The σ bands and Fermi surfaces in the electronic band structure of $MgB_2$ contain approximately cosine shaped bands and Fermi surface cross-sections that are associated with TB equations. Hopping integrals, also known as transfer coefficients, are essential parameters in TB equations. Many hopping models have been created to explain superconductivity, particularly for the high $T_c$ cuprates.[24, 25] Such models become more powerful when informed by parameters extracted from DFT calculations, since these calculations deal with collective charges, or quasiparticles, instead of electrons. In this context, DFT provides a direct link between crystal structure and material properties.[26]

In this work, we focus on $MgB_2$ as a case study for charge transfer, or “hopping”, mechanisms using *ab initio* DFT methodology described in detail in earlier publications.[4, 5, 20] The same methodology has been used to interrogate band structures and Fermi surfaces of superconductors with different compositions and attributed superconductivity mechanisms, such as for $CaC_6$, $LaH_{10}$ and β-$FeSe_{1-x}$ with equal success.[7-10] This article adds further detail to our earlier DFT calculations on EBSs for $MgB_2$ which describe cosine shaped bands that support the modification of TB equations to identify a superconducting gap.[4] By addressing band structures near $E_F$, yet away from principal high symmetry nodes, we show that these bands, Fermi surfaces and topologies illustrate a hopping mechanism for superconductivity.

## 2. Results

The following examples show conventional high symmetry directions for the EBS of $MgB_2$ as well as directions offset from high symmetry nodes within the Brillouin zone. With this comparison, we address band degeneracies at, or near the Fermi surface, $E_F$. Other examples are further discussed in Section 3 and Supporting Information.

### 2.1. Band structures at 0 GPa

Based on earlier calculations and experimental data,[4, 5, 20] we utilise a superlattice, doubled along the *c*-axis direction with space group $P\overline{6}c2$, to elucidate key features of the $MgB_2$ band structure. **Figure 1a** shows the electronic band structure for $MgB_2$ at 0 GPa calculated along high symmetry reciprocal directions using the GGA functional. The Γ–A region (red outline) is the focus of this work. Reciprocal directions for the 2*c* superlattice are shown in **Figure 1b**.

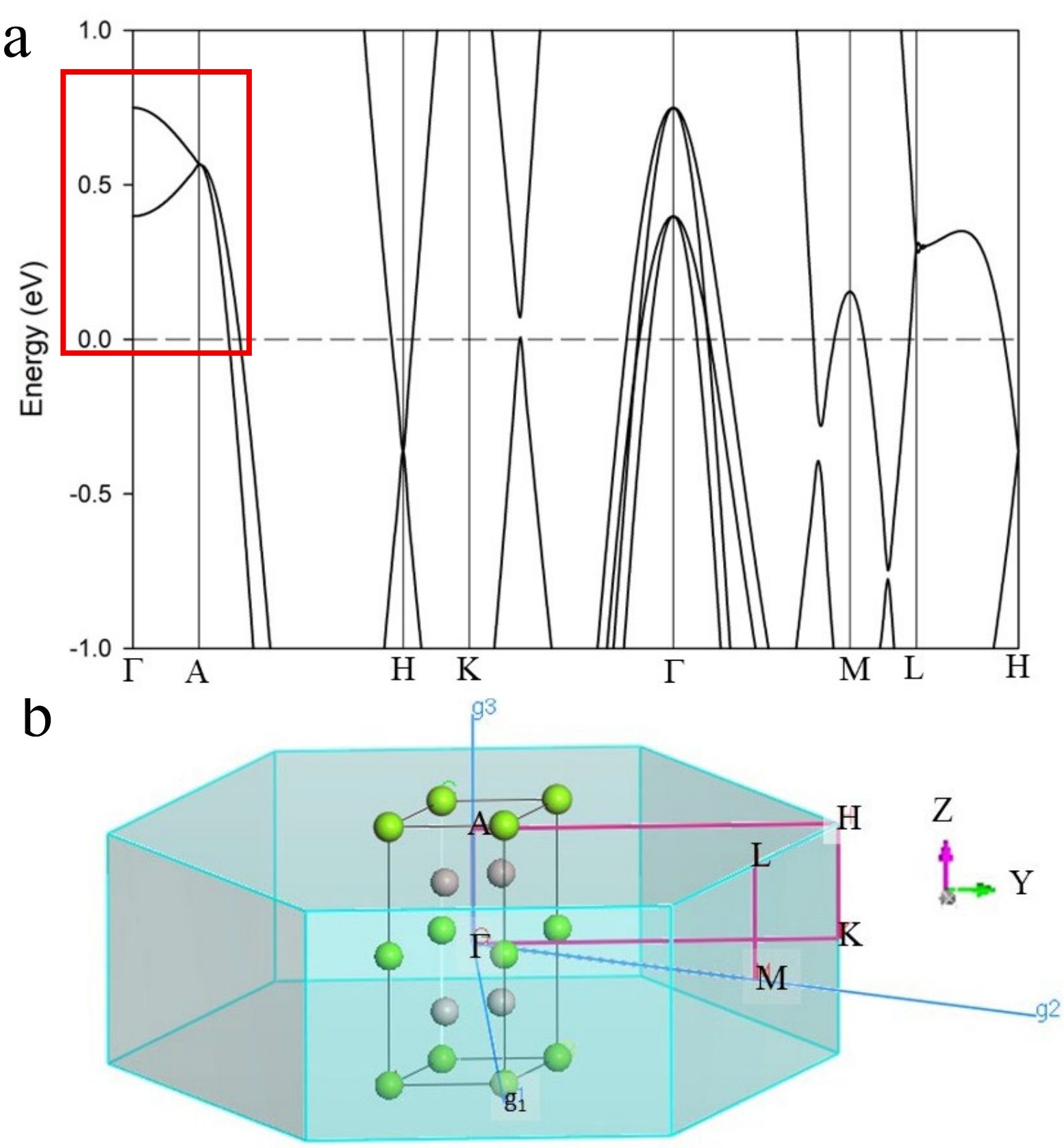

**Figure 1.** (a) Electronic band structure for the 2*c* superlattice of $MgB_2$ at 0 GPa calculated along high symmetry directions (SG: $P\overline{6}c2$). The Γ–A region (red outline) is the focus of this work. (b) reciprocal space orientations for the 2*c* $MgB_2$ superlattice.

**Figure 2** shows key segments of the EBS for the $MgB_2$ superlattice at Γ–A and Γ–M as well as for directions parallel to Γ–A at regular intervals along the Γ–M direction. These segments sample the band structure at intervals within the Brillouin zone that approximate nodal inflection points for light and heavy effective mass σ bands. The approximate loci of inflection points, between (0, $k_y$, 0) and (0, $k_y$, 0.5) are obtained from calculations previously published for $k_F^{H,L}$.[4, 5] The value and derivation of $\Delta E_{gap}$ (at 7.9 meV) shown for the Γ–A section in Figure 2a is described in detail in an earlier publication.[4] At $k_y$ = 0.097, the degenerate bands along Γ–A split into two separate bands that cross near $E_F$ with an overlap at Γ of 29.0 meV. For this calculation series, the overlap at Γ reduces to 1.3 meV within ± 1.0 meV of $E_F$ for $k_y$ = 0.101. Similar calculations undertaken using the GGA functional are shown in Supporting Information (**Figure S1**).

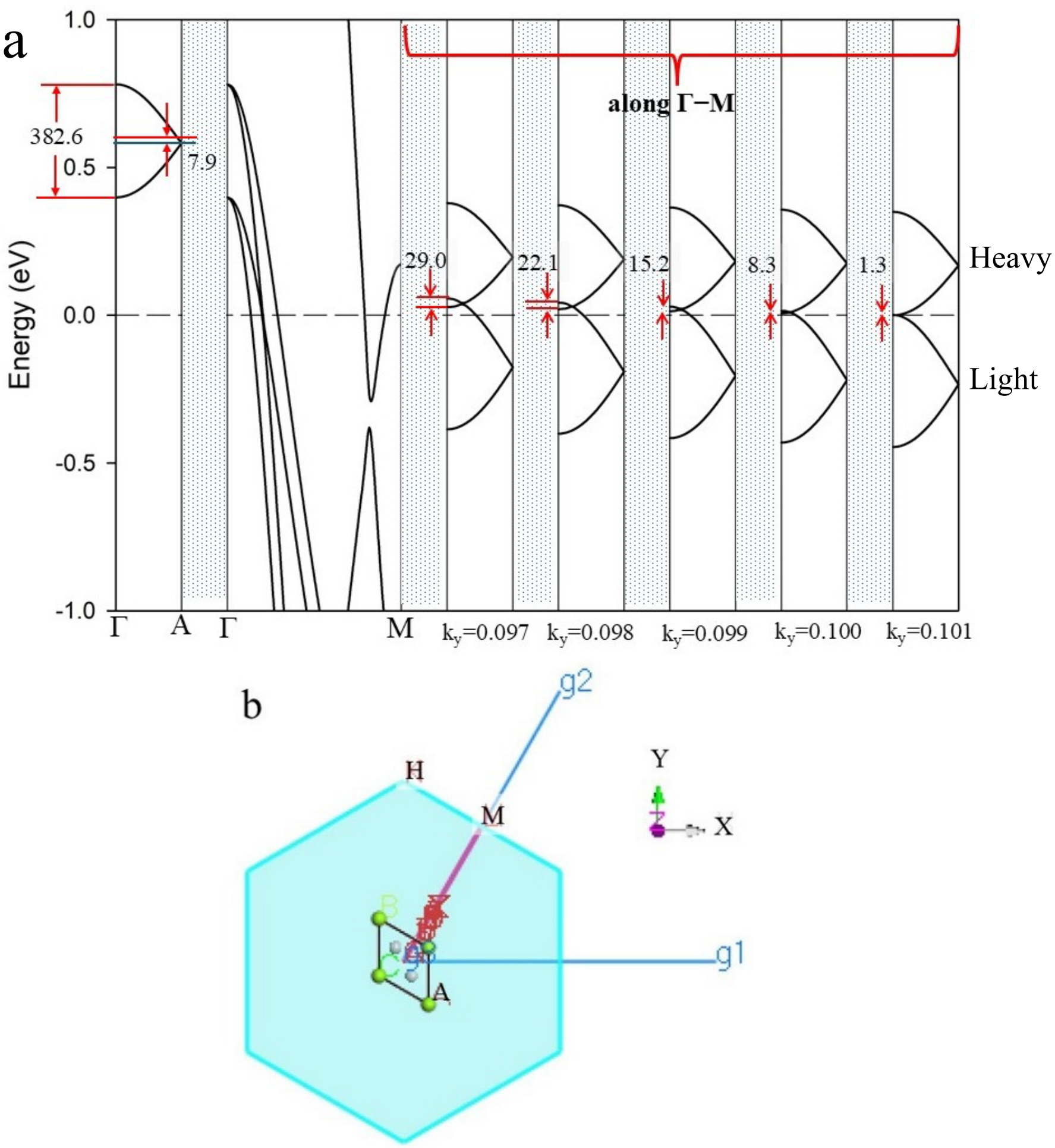

**Figure 2.** (a) EBS of a 2*c* superlattice for $MgB_2$ at 0 GPa calculated using the LDA functional along reciprocal directions parallel to Γ–A direction and intersecting along Γ–M at approximate nodal inflection points for the light and heavy effective mass σ bands (i.e. at (0, $k_y$, 0) and (0, $k_y$, 0.5)). When the folded light and heavy effective mass cosine bands meet at the Fermi level,

the corresponding $k_y$ value is the component of the common Fermi vector. Energy values for bands at Γ–A and Γ–M are in meV; values for $k_y$ are shown at bottom of plot for each segment along Γ–M. (b) reciprocal space orientations for the 2*c* superlattice of $MgB_2$ showing the $k_y$ intersections along Γ–M used to generate data points in (a).

2.1.1. Corresponding Fermi surfaces

**Figure 3a** shows Fermi surfaces for the light and heavy effective mass quasiparticles of σ bands in the 2*c* superlattice for $MgB_2$ at 0 GPa. The region between the inner tubular sections of the Fermi surface (parallel to Γ–A) and the outer shaded section highlight the extent and shape of these warped cylinders[27] of σ bands. At zero temperature, these regions represent states populated with electrons or with empty states for the light and heavy effective masses of the same **k** vector, respectively.

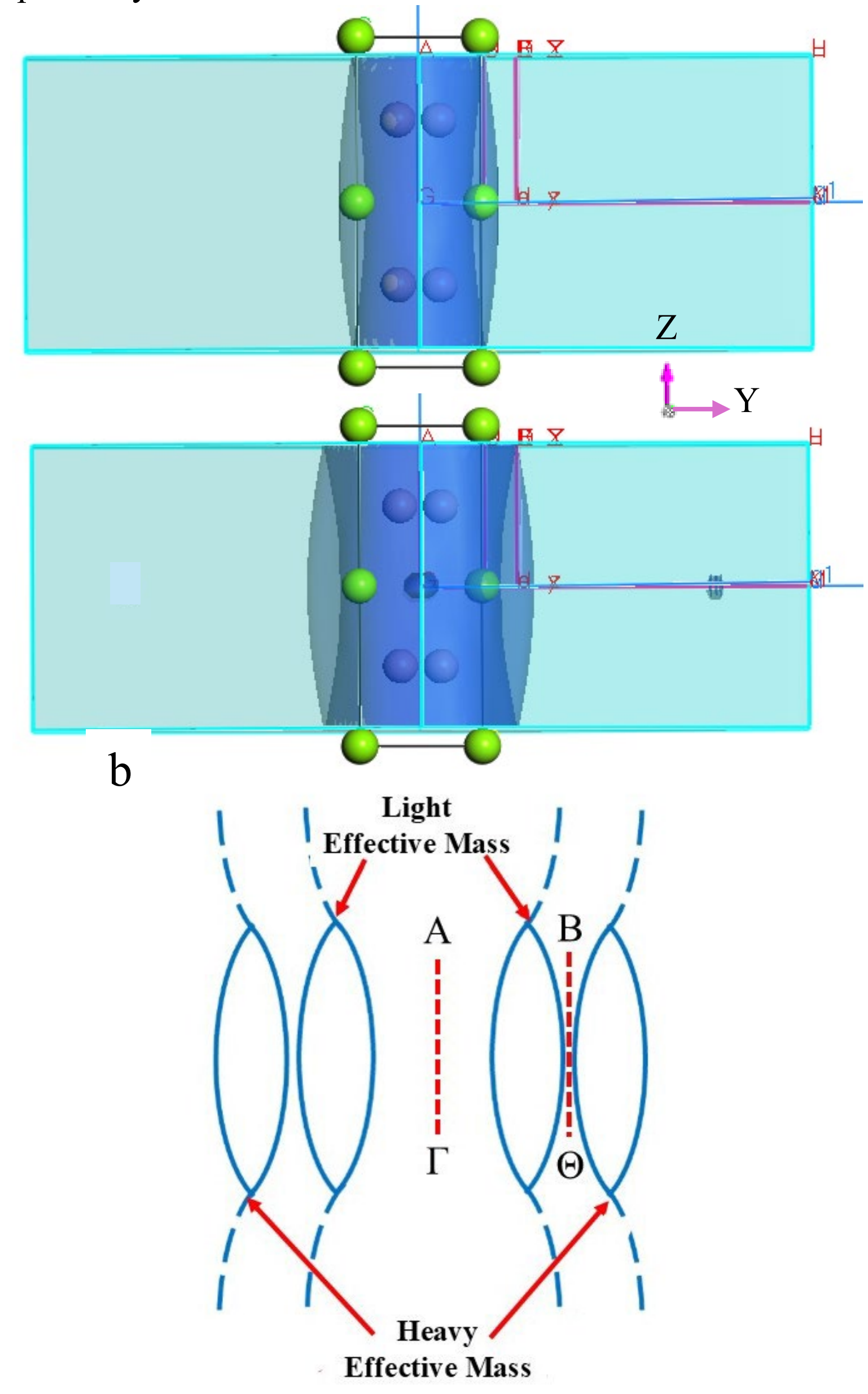

**Figure 3.** (a) Fermi surfaces for the light and heavy effective mass σ bands of the 2*c* superlattice for $MgB_2$ at 0 GPa. (b) Schematic of the projected warped tubular Fermi surfaces for a single $MgB_2$ unit cell and the folded bands of a 2*c* superlattice. Reciprocal directions for Γ–A and band segments (Θ–B; $0.091 < k_y < 0.101$) shown in Figure 2 are shown as dotted red lines.

**Figure 3b** is a schematic highlighting the origin of Fermi surfaces for the 2*c* superlattice of $MgB_2$ by folding the single primitive cell Fermi surface. The band segments displayed in Figure 2a (i.e. $0.097 < k_y < 0.101$) have been calculated along a reciprocal direction that approximately lies between the two lens shaped projections of the folded cosine bands in Figure 3b. Note that the folded light and heavy effective mass Fermi surfaces do not intersect at 0 GPa.

2.2. Band structures at high pressure

**Figure 4** shows the electronic band structures of the 2*c* superlattice for $MgB_2$ at 4 GPa and 8 GPa, parallel to the Γ–A axis (at regular intervals along Γ–M) calculated using the LDA functional. Calculations for these samplings along Γ–M follow the same procedure as indicated for Figure 2a. Similar to the case for 0 GPa, the separated light and heavy σ bands cross in proximity to the Fermi level with greater overlap at Γ compared to that in Figure 2a. At 8 GPa (Figure 4b), this trend for band splitting near $E_F$ continues with greater overlap of the light and heavy bands at Γ. At 4 GPa and 8 GPa, the Fermi level intersects the lower band at $k_y = 0.097$ and at $k_y = 0.101$ intersects the higher energy band.

Supplementary **Figure S1**, calculated for the 2*c* superlattice of $MgB_2$ using the GGA functional along the same directions, shows that degenerate bands at Γ–A also split into two bands that cross near $E_F$. For the GGA calculation, bands at Γ follow a different trend to that shown in Figure 2 and Figure 4 in that the "overlap" is minimal along Γ–M closer to Γ and leads to a separation of bands close to $E_F$ at higher $k_y$ values.

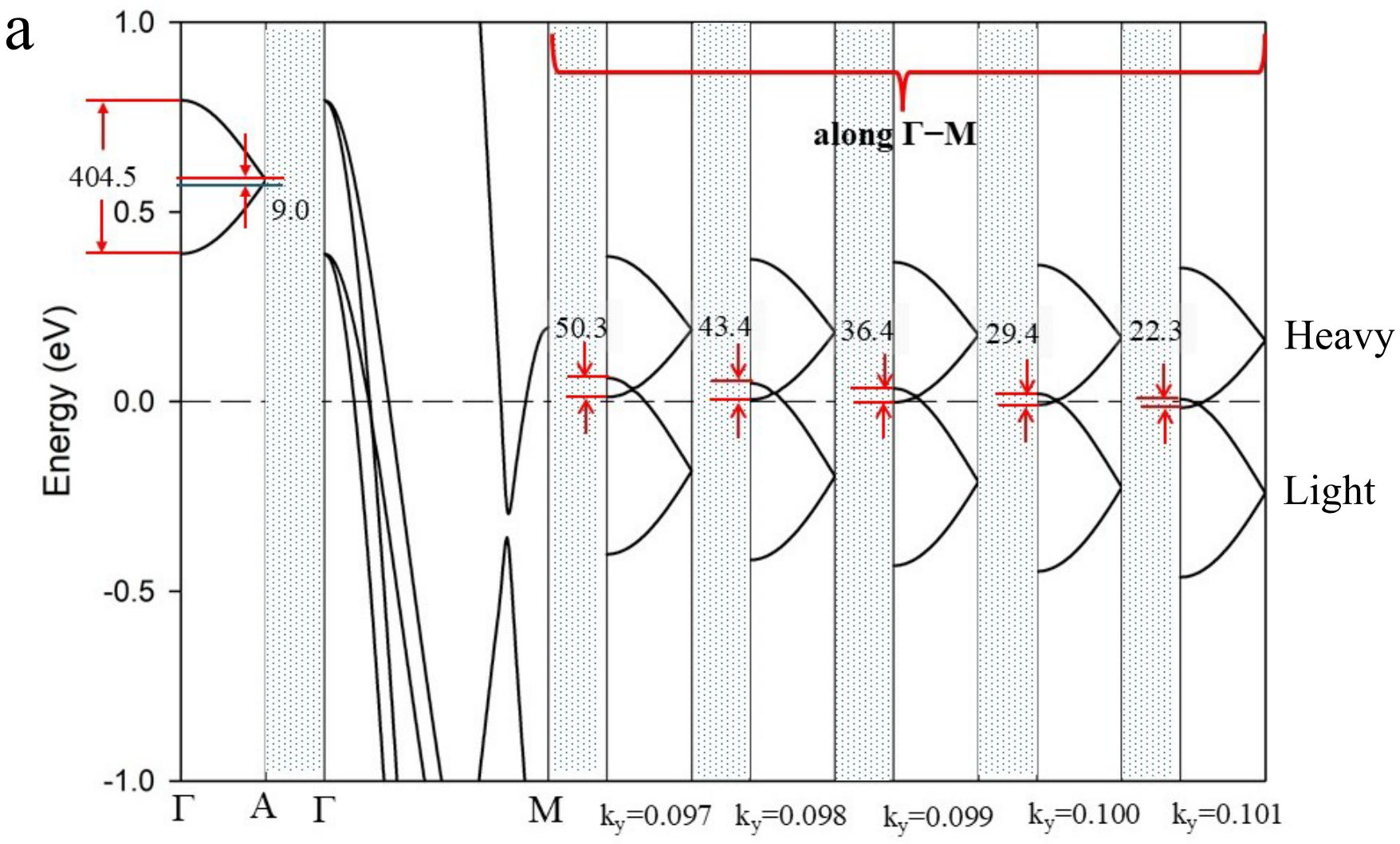

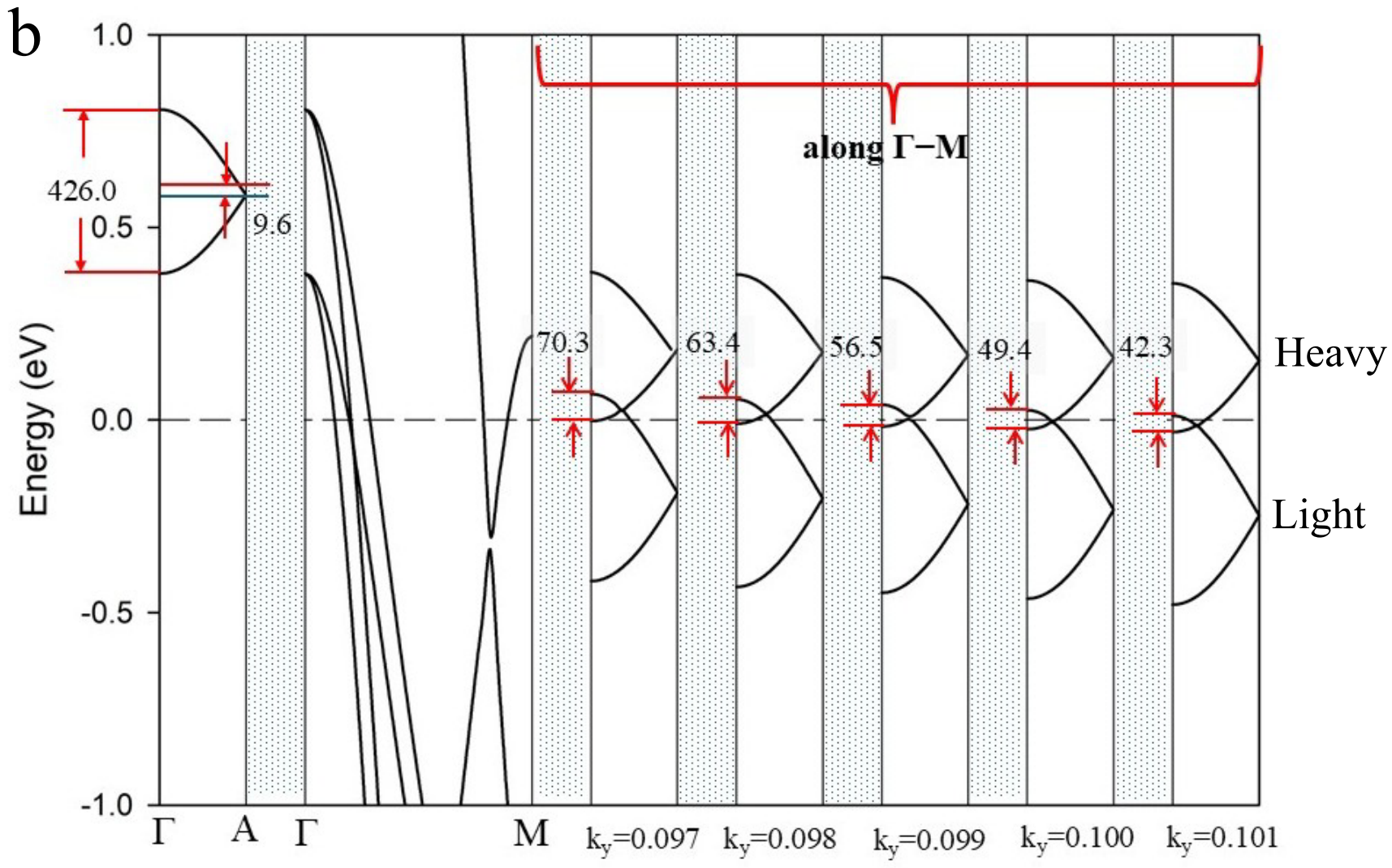

**Figure 4.** EBS for a 2*c* superlattice of $MgB_2$ calculated using the LDA functional at (a) 4 GPa and (b) 8 GPa. Bands are calculated along lines that are between the light and heavy effective mass σ band Fermi surfaces, as shown in Figure 3. These bands refer to the distance from the origin (0, 0, 0) to the Brillouin zone boundary (at $k_y = 0.5$). When the folded light and heavy effective mass cosine bands meet at the Fermi level, the corresponding $k_y$ value is the component of the common Fermi vector. Energy values for bands at Γ–A and Γ–M are meV; values for $k_y$ are shown at bottom of figure for each segment along Γ–M.

For completeness, **Figure S2** (Supporting Information) shows band structures at segments along Γ–M for values of $k_y < 0.091$ and $k_y > 0.101$ in comparison to those used in Figure 2, Figure 4 and Figure S1. For $k_y < 0.091$, the degenerate bands along Γ–A split into two bands with intersection above $E_F$. For ky >0.101, the bands are separate with no intersection or overlap at Γ. For $k_y \geq 0.124$, only the heavy effective mass bands intersect with $E_F$. These calculations show that the Fermi surface intersects the light effective mass bands for $k_y < 0.097$. For $k_y >$ 0.101, the Fermi surface intersects the heavy effective mass band.

2.2.1. Corresponding Fermi surfaces at high pressure

**Figure 5** shows the light and heavy effective mass Fermi surfaces for σ bands of the 2*c* superlattice for $MgB_2$ at 4GPa and 8GPa. Note that the folded light and heavy effective mass Fermi surfaces do intersect at 4 GPa and 8 GPa. At 8 GPa, an extended flat section on the outer side of the light effective mass Fermi surface develops suggesting that the inner side of the heavy effective mass Fermi surface (see region next to the bracket in Figure 5b) may intersect with the light effective mass at higher pressure.

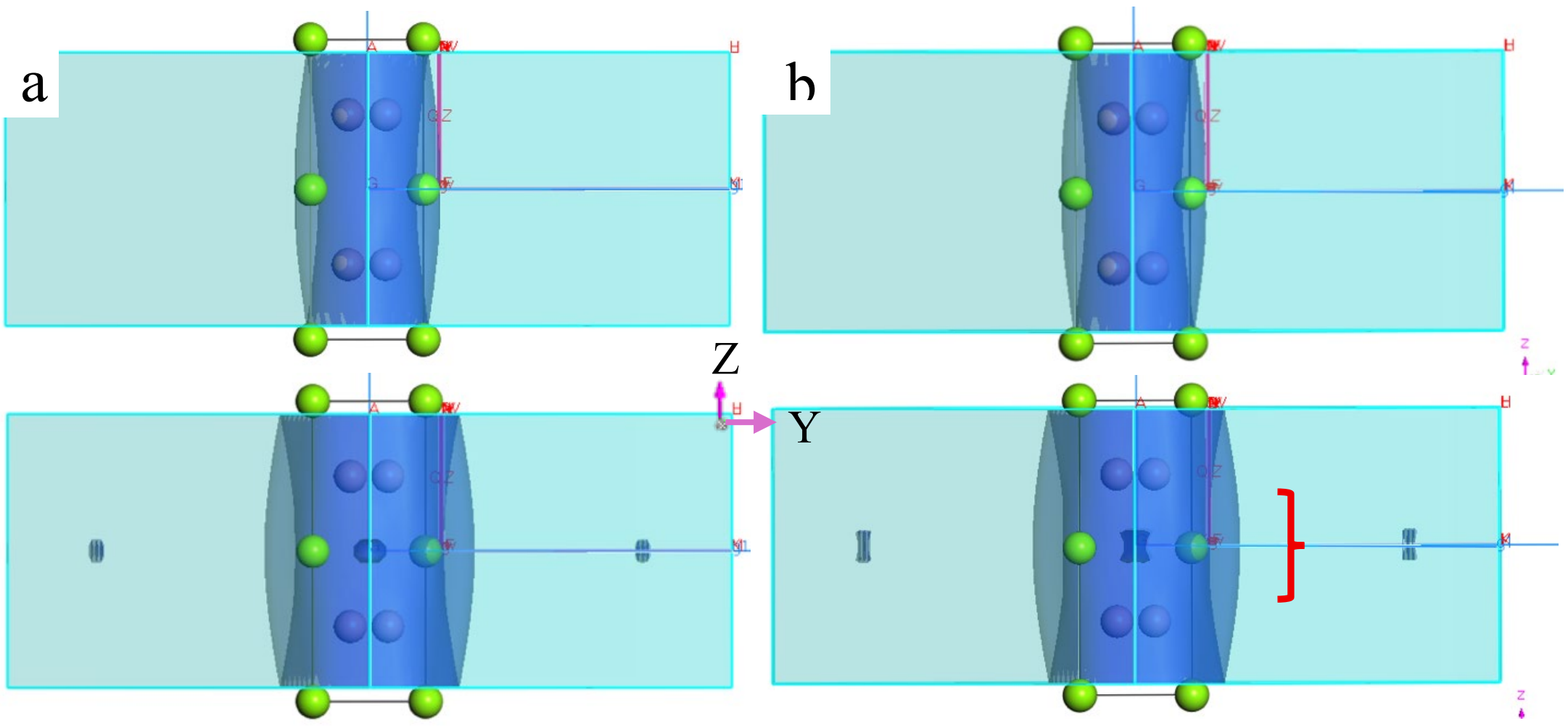

**Figure 5.** Fermi surfaces for the light (top) and heavy (bottom) effective mass σ bands of the 2$c$ superlattice for $MgB_2$ (a) at 4 GPa and (b) at 8 GPa.

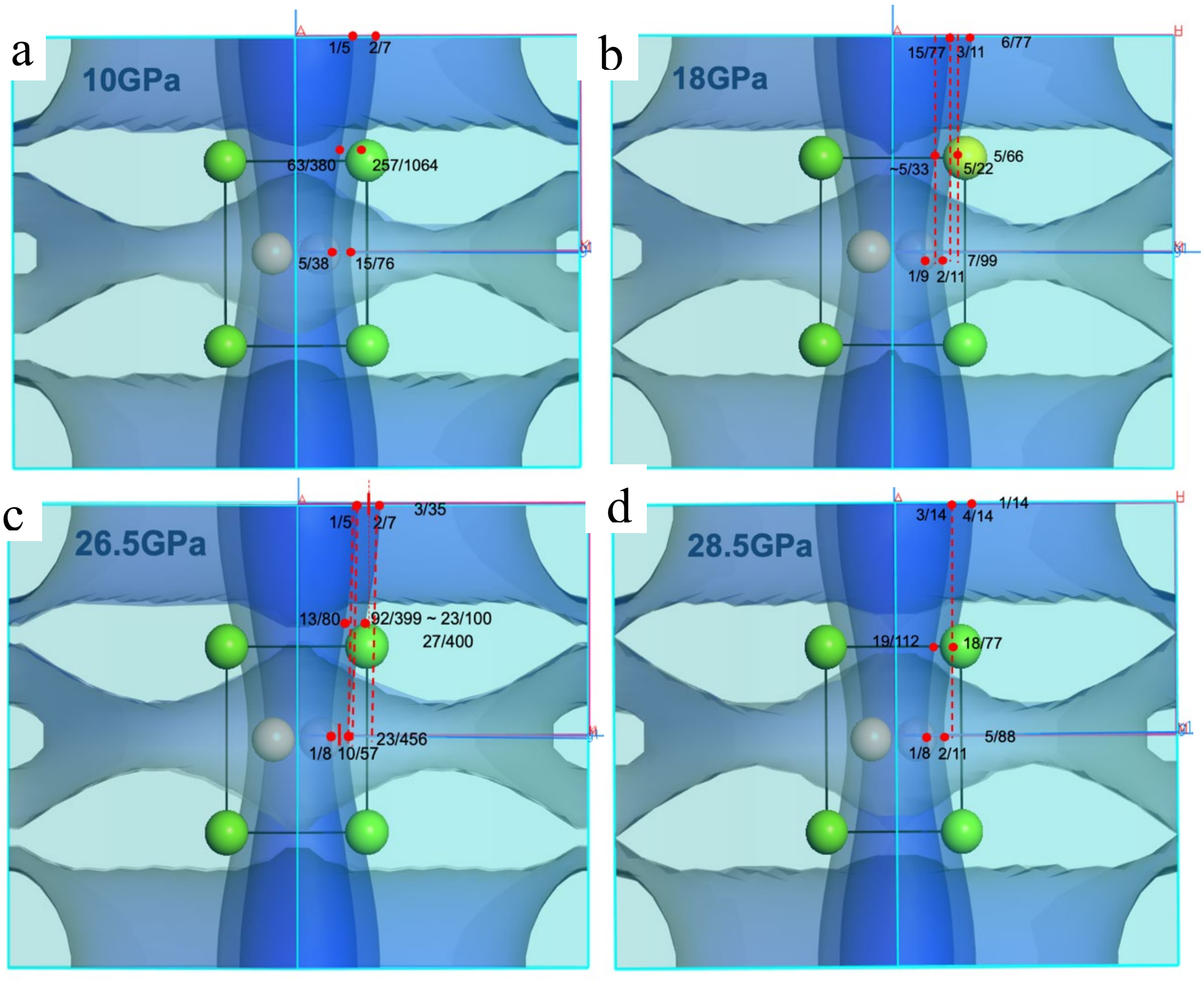

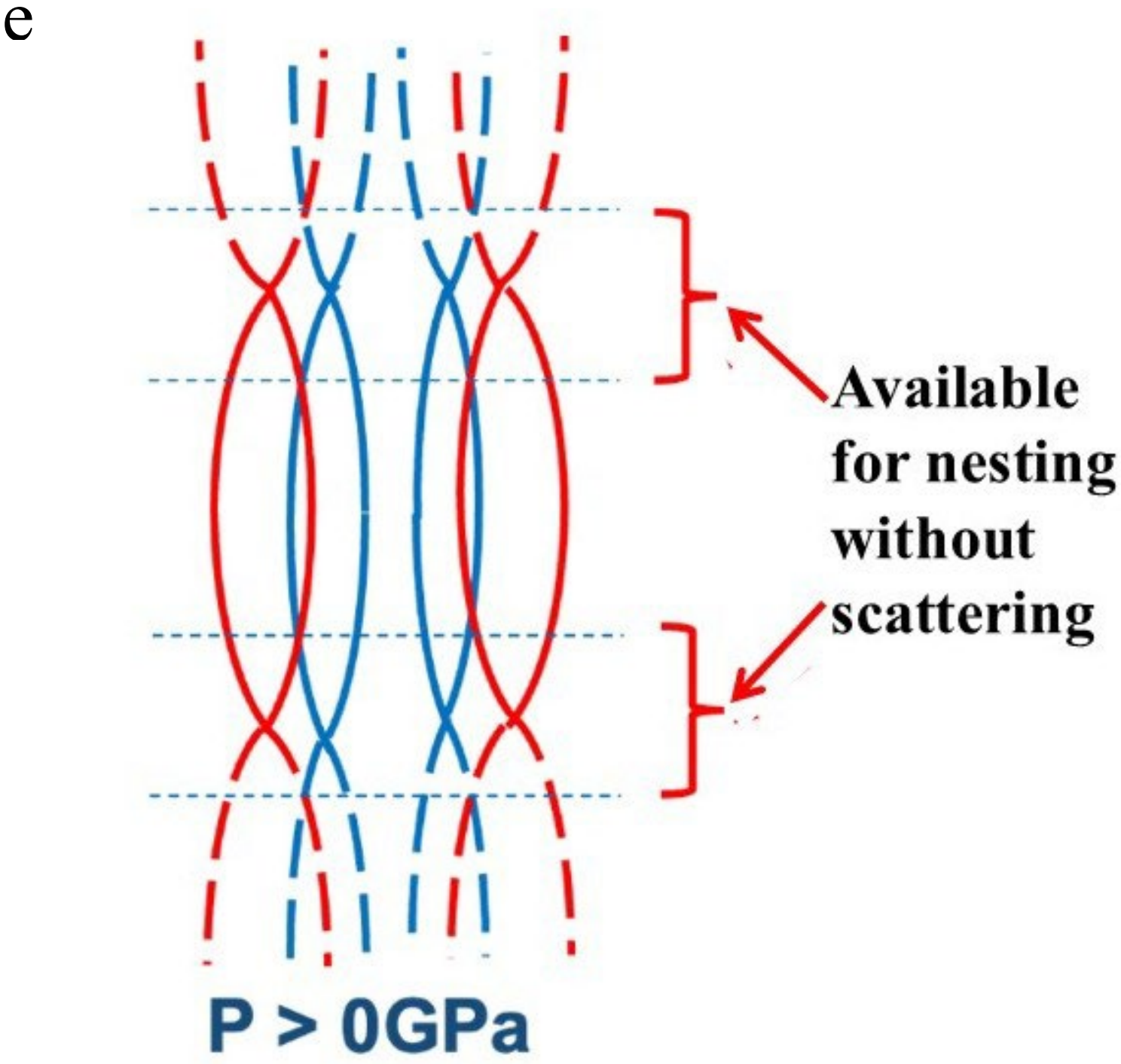

**Figure 6.** (a) Fermi surfaces for the light and heavy effective mass σ bands of the $MgB_2$ 1*c* primitive cell at (a) 10 GPa, (b) 18 GPa, (c) 26.5 and (d) 28.5 GPa. Figure 6e shows a schematic of intersecting Fermi sections that occur upon folding at the inflection point with elevated pressure. Approximate coordinates of key intersection points (at the centre, inflection and Brillouin zone boundary points) of the projected Fermi surface are displayed as fractions of Γ–M.

2.2.2. Fermi surfaces at other pressures

**Figure 6** shows Fermi surfaces for the 1*c* primitive cell of $MgB_2$ with increasing applied pressure. These Fermi surface calculations are used to graphically extract approximate coordinates for key points that enable analysis of band amplitudes (e.g. at the centre, inflection and Brillouin zone boundary points). The magnitudes of geometrical projections suggest that the folded Fermi surfaces have intersections. These geometrical values and intersections are discussed in more detail below and in Supporting Information.

3. Discussion

As is well known, electrons at the Fermi level in a narrow window of energies of order the thermal energy determine the electronic transport properties of solids.[28, 29] For electrons with energy well below the Fermi level, significantly higher energies are required to excite these electrons into available empty electronic states. This requirement, based on Hund's rules and Pauli's principle, also applies to lowest energy empty states near the Fermi level. The interplay of electrons moving between filled and empty electronic states within a solid connected to two external electrodes facilitates electron transport between the electrodes. For this reason, the electron energies that are involved in transport phenomena belong to the narrow window of energy at the step transition of the Fermi-Dirac distribution (i.e. between fully populated and fully unpopulated electronic states below and above the Fermi level, respectively).

The introduction of a bandgap at the Fermi level can lower the energy of a metal.[6, 30-32] This generally happens in two typical ways; one leads to formation of a charge density wave (CDW) state and the other, to formation of a superconducting state.[30] In both cases, the lowering of energy involves interaction of occupied states with unoccupied states lying close to the Fermi level. However, these states differ in the way they interact.

A metal-to-insulator phase transition takes place during formation of a CDW, while a metal-to-superconductor phase transition is induced in the other case. Interband pair-state excitations become favourable when two Fermi surfaces are very similar, or nested, and when the metal has a large average phonon frequency $\langle \omega \rangle$.[30] If more than one partially-filled band is present, a superconductor with $T_c$ well above the McMillan limit is often explained as due to a "multiband" mechanism, while remaining within the confines of the BCS theory for superconductivity.[30] DFT calculations provide reliable Kohn-Sham one-electron band energies and wavefunctions.[33] From this viewpoint of one-electron band theory, a superconducting state also involves orbital mixing among band levels above and below the Fermi level.[18]

Typically, DFT calculations of electronic band structures are undertaken and displayed along high symmetry directions. In general, such two-dimensional representations are suitable measures of atomic-scale and bulk material properties based on characteristics and format of the displayed bands. Results shown in Figures 2 and 4 show that band characteristics for sections in close proximity to the Fermi surface (i.e. along directions parallel to Γ–A at regular intervals along Γ–M) can differ from that of bands along high symmetry directions.

The presence, or otherwise, of band degeneracy is a case to consider. For example, degenerate bands occur along Γ–A, but do not occur as degenerate electronic bands along reciprocal directions closer to the Fermi surface where electronic transport occurs. For the sections displayed in Figures 2 and 4 (along the Γ–M direction), degenerate bands only occur at the nodal point "A" for all pressures. Thus, degenerate electronic bands, generally expected to significantly impact physical properties of solids, may not always occur in close proximity to the Fermi surface. Band segments shown in Figure 2 and Figure 4 confirm that DFT effectively accounts for true topology, degeneracy and band crossings in close proximity of the Fermi surface, all of which influence material properties.[26, 33]

3.1 Intra-band and inter-band hopping

As known from textbooks[17, 19, 34] and demonstrated in earlier publications[4, 7, 9], bands for a $2c$ superlattice can be geometrically constructed by folding bands that connect high symmetry nodes at the mid-point. This folding results in lens shaped cross-sections as shown in Figures 2 and 4. High resolution DFT calculations have also shown that the bonding and antibonding sections of cosine shaped bands show asymmetry.[4] This cosine band asymmetry, measured as a difference in energy, correlates well with the magnitude of the superconducting gap[4, 7-9] and is mathematically described using corrected TB equations (1):

$$E^{H,L}\left(k_x, k_y, k_z\right) = E_0 - (2t_\perp - \Delta)\cos(ck_z) - \frac{\hbar^2}{2m^{H,L}}\left({k_x}^2 + {k_y}^2\right) \qquad 0 \leq ck_z \leq {}^{\pi}/_{2}$$

$$(1)$$

$$E^{H,L}\left(k_x, k_y, k_z\right) = E_0 - (2t_\perp + \Delta)\cos(ck_z) - \frac{\hbar^2}{2m^{H,L}}\left({k_x}^2 + {k_y}^2\right) \qquad {}^{\pi}/_{2} \leq ck_z \leq \pi$$

The degeneracy of the electronic bands along the Γ–A direction ($k_x, k_y = 0$), while not representative of electronic behaviour at the Fermi surface (as noted above), conveniently allows for identification of coincident TB parameters in equation (1). Similarly, evaluation of equation (1) on the respective light and heavy mass Fermi surfaces at their inflection points (i.e. $k_z = {}^{\pi}/_{2c}$ based on the $1c$ primitive unit cell) allows identification of energy equivalences at those points:

$$E_F = E\left(k_{Fx}^{H,L}, k_{Fy}^{H,L}, {}^{\pi}/_{2c}\right) = E_0 - \frac{\hbar^2}{2m^{H,L}}\left({k_{Fx}^{H,L}}^2 + {k_{Fy}^{H,L}}^2\right)$$

$$(2)$$

$$E_0 = E_F + \frac{\hbar^2}{2m^H}\left({k_{Fx}^H}^2 + {k_{Fy}^H}^2\right) = E_F + \frac{\hbar^2}{2m^L}\left({k_{Fx}^L}^2 + {k_{Fy}^L}^2\right)$$

Thus, even though the high symmetry direction Γ–A is not directly representative of electrons at the Fermi surface, convenient access to important coefficients of the TB equations is provided. In addition, the relations between light and heavy effective mass band coefficients can be determined. The equations at $k_x, k_y = 0$ also show that the Δ correction in equation (1) modulates a perfectly symmetric TB dependence with a cosine function.

3.2 Coupling of electron-hole pairs

Cosine bands that cross $E_F$ and join high symmetry nodes, combined with an absence of degeneracy for bands in proximity to the Fermi surface, provide further hints on a superconductivity mechanism for $MgB_2$. Analyses clarifying the conditions for modified TB equations based on cosine band asymmetry have been discussed previously in the context of Bloch orbitals for an infinite linear chain with two s-states.[17, 19, 30, 34] For the sake of clarity, we summarise key concepts below.

Two s-states per atom are considered hybridised on each atom, that is, the atomic contribution to the Bloch state for the linear chain is treated as a linear combination of the two atomic states. To construct the Bloch state for the linear chain, each hybridised atomic state is then multiplied by the exponential factor $e^{ikma}$ and summed for all atoms in the linear chain. Crystal orbitals can also be constructed as linear combinations of the Bloch orbitals[17, 18]

With use of the Schrödinger equation and projection onto each atomic state leads to a secular determinant for non-trivial solutions. Evaluation of the expression along the Γ–A direction $(k_x, k_y = 0)$ takes the form:

$$\begin{vmatrix} E_0 - 2t_\perp \cos(ck_z) - E_k^H & \Delta Cos(ck_z) \\ \Delta Cos(ck_z) & E_0 - 2t_\perp \cos(ck_z) - E_k^L \end{vmatrix} = 0 \qquad (3)$$

Equation (3) can be solved to obtain the $k_z$ components of the equations shown in (1). The resulting bands are schematically displayed for a $1c$ primitive cell and an equivalent $2c$ superlattice in **Figures 7a and 7b**, respectively (compare Figure 7a to Figure 5.2 in[34]).

The terms $2t_\perp \cos(ck_z)$ and $\Delta cos(ck_z)$ correspond to 'onsite' matrix elements and hopping integrals between adjacent atoms, respectively. An intra-band excitation, when using a cosine shaped band for the $1c$ primitive unit cell of $MgB_2$, becomes an inter-band excitation when using a folded cosine band for the $2c$ superlattice. Note that in the schematic of Figure 7, a cosine shaped band represents the cross-section of the Fermi surfaces shown in Figures 2 and 4.

Alignment of a nodal inflection point with the Fermi level occurs at axes parallel to the Γ–A direction that intersect the Fermi surface at $k_z = {}^{\pi}/_{2c}$, as shown in **Figure 7c** and Figure 2. This alignment, which closely resembles the electron and hole properties at, or in close vicinity to, the Fermi surface clearly indicates that a bonding-antibonding coupling between electrons and holes, can be activated. This concept is consistent with the picture of nesting identified from analysis of the topology of bands and Fermi surfaces when superlattices are used for calculations as shown in earlier work.[4] Such nesting has been identified for sections of separate light and heavy effective mass σ Fermi surfaces of $MgB_2$.[4] Since these nested regions are joined by respective (intra-) light and (intra-) heavy effective mass nesting vectors, the difference between these nesting vectors will also define inter-light and inter-heavy σ Fermi surface nestings. Alignment of a nodal inflection point with $E_F$ is equivalent to electron-hole pair coupling and hybridisation commensurate with gap formation at the nodal point, as previously described in the literature[18].

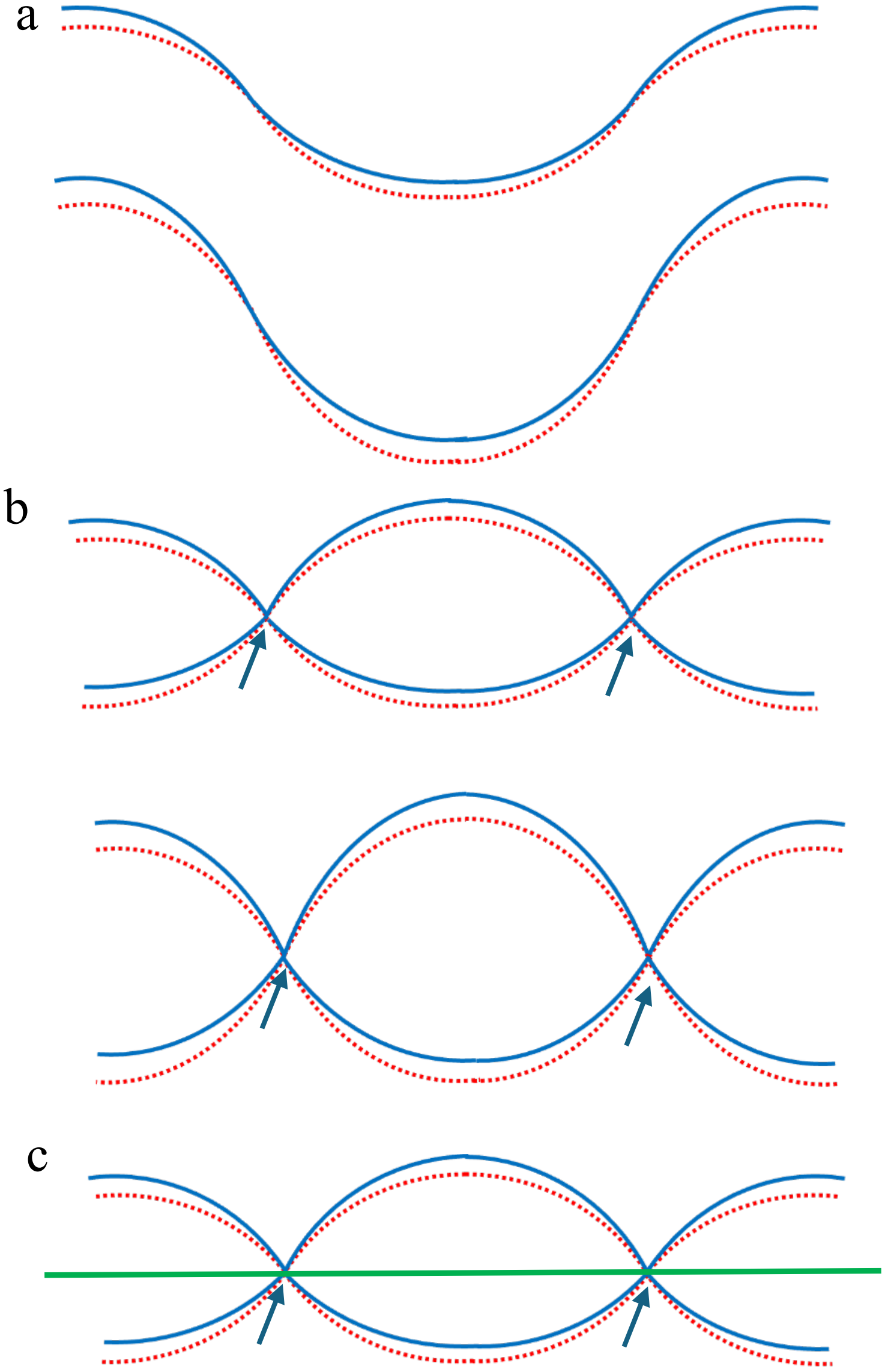

**Figure 7.** Schematic of cosine shaped $MgB_2$ electronic bands for (a) a 1$c$ primitive unit cell, (b) a 2$c$ superlattice, and (c) a 2$c$ superlattice for bands aligned with the Fermi level (green line). The red dotted curves represent an electronic band in the absence of a hopping modulation of Δ amplitude between the two different s states in adjacent atoms. The blue continuous curves include hopping between the two different states in adjacent atoms. Note that the red dotted and blue continuous curves have the same energy at the inflection mid-point $k_z = {}^{\pi}/_{2c}$ in Figure 7a. This aspect of band character becomes apparent in the folded representation shown in Figure 7b, where the point $k_z = {}^{\pi}/_{2c}$ (arrowed) becomes a nodal point.

We show that DFT calculations undertaken at high resolution carry key details, including mechanisms of electronic transport because DFT calculations embrace quasiparticle distributions.[26] For example, when the geometry of a crystal is slightly modified during calculation, the electron density undergoes a redistribution to optimise, or minimize, energy. Therefore, responses to small (e.g. sub-micrometer or sub-picometer), geometric changes are collective rather than individual or single electron. An attribute of the LDA functional is that the accuracy of bond lengths and geometries of solids is ~ 1%.[26] As such, DFT appears to be more appropriately associated with 'quasiparticle responses' rather than individual electron behaviours. At conclusion of a converged geometry optimisation process, where minute geometry changes are required for convergence and the bands below the Fermi energy are nearly filled, we envisage that a redistribution of the last electron charge leaves behind a hole that the collective sea of adjacent electrons must accommodate.

### 3.3 Intersecting Fermi surfaces and band coherence

When DFT calculations show crossing of bands and Fermi surfaces with different slopes at the intersection, avenues or 'pathways' become available for electrons to change velocity or momentum. Such an intersection enables scattering and loss of potential coherency. Coherent nesting has been previously identified for the folded Fermi surfaces of $MgB_2$, particularly at, and in the vicinity of, inflection points.[4] These nesting regions become apparent when using 2*c* superlattices for DFT calculations but may remain unnoticed if working on a 1*c* primitive cell or higher symmetry configuration (e.g. P6/mmm). We note that the use of a 2*c* superlattice for DFT calculations of $MgB_2$ is inspired by experimental observation of superlattice peaks using Raman and synchrotron THz spectroscopies.[5, 20]

The thermal energy carried by a collection of nearly free electrons is proportional to the number of electrons times the average thermal energy contributed through their degrees of freedom.[28, 29] We posit that the volume fraction of Fermi surface where coherent nesting exists, before an intersection occurs (i.e. a different band slope), will be proportional to the superconducting temperature. In other words, superconductivity is associated with the thermal energy that can be added to a system of coherent electrons before such coherency is destroyed.

The approximate position(s) of band crossing or intersection at various selected pressures can be calculated using relative distances extracted from graphical outputs for calculated Fermi surfaces. These positions have been estimated graphically for 10 GPa, 18 GPa, 26.5 GPa and 28.5 GPa using the calculated Fermi surfaces and referring to distances between key points labelled in Figure 6. Further details on the calculations of key reciprocal distances are given in Supporting Information (**Table S1**). These calculations result in respective fractions of 0.83457, 0.6688, 0.5206 and 0.4621 for the total (folded) $k_z$ directions that can be nested before band intersections destroy coherency. By comparing these fractions of coherent nesting at pressure to that for $MgB_2$ at 0 GPa (i.e. a Fermi surface without band intersections), estimates for $T_c$ at 10 GPa, 18 GPa, 26.5 GPa and 28.5 GPa, are 32.5 K, 26 K, 20.3 K and 18 K, respectively.

For reference, these estimated $T_c$ values based on coherent nesting are shown in **Figure 8.** Figure 8 compares experimentally determined $T_c$ values as a function of pressure[35, 36] and an alternative method to calculate $T_c$ using the thermal energy of the Kohn phonon anomaly.[37] The estimated $T_c$ values determined by this alternative DFT-based method are comparable to experimental values and to that based on the $E_{2g}$ phonon.[37, 38] Note that the graphical estimates for $T_c$ at pressure in Figure 8 are less precise than some numerical methods because additional indirect steps *via* image contrast and image resolution have been used instead of exact equations.[39] The accuracy of $T_c$ estimates is improved at higher pressures because the intersections occur closer to inflection (nodal) points of folded cosine shaped bands (where lower curvature is apparent). Higher mathematical accuracy could be achieved by using a function of the cosine dependence. However, the improvement is marginal and does not add significant value to the proof of this approach.

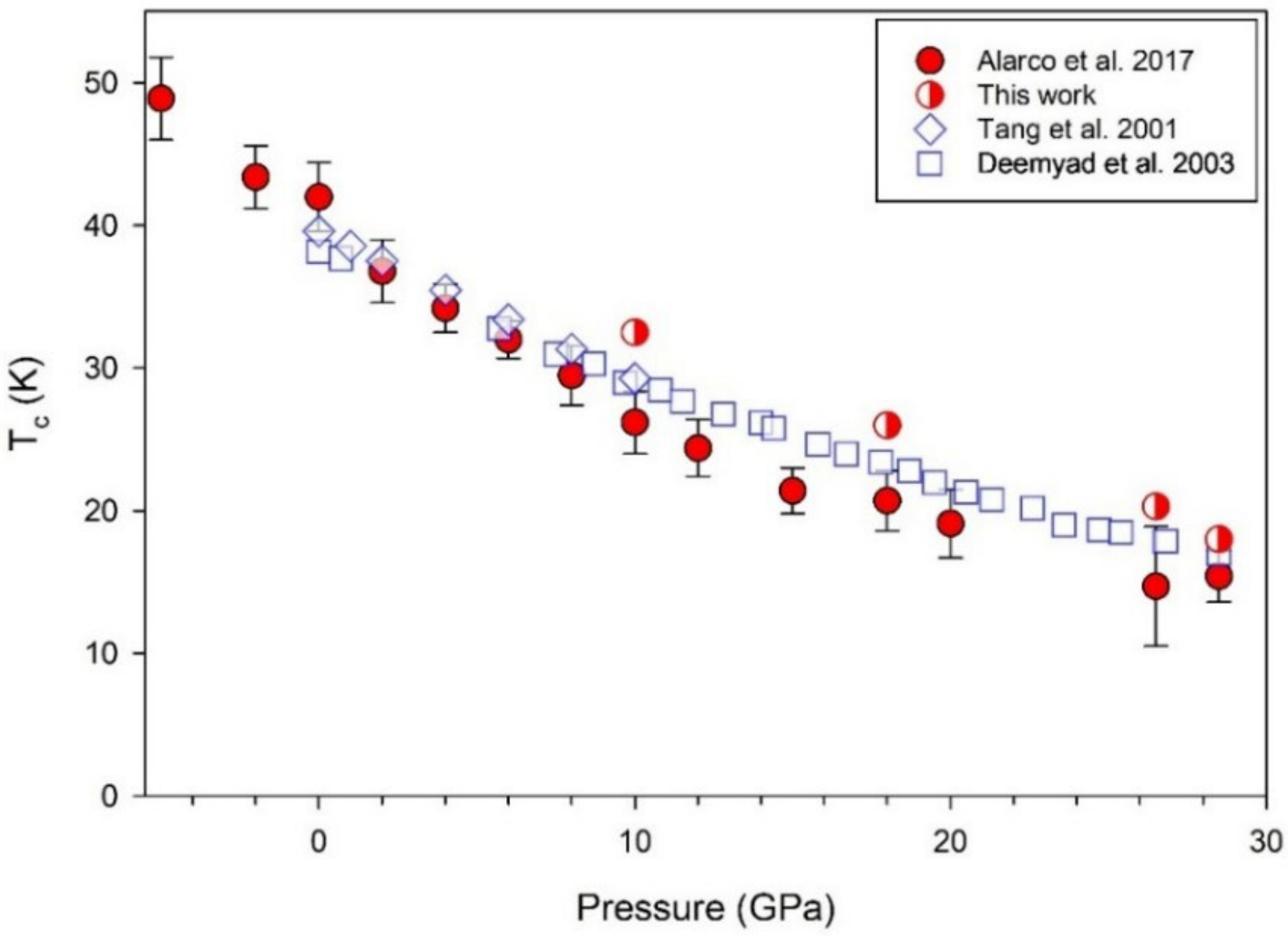

**Figure 8.** Estimated $T_c$ based on calculated amplitudes of asymmetry for the cosine shaped band at the band intersection (half-filled symbols), compared with experimentally determined superconducting transition temperatures (open symbols)[35, 36] and calculated values using changes to the Kohn anomaly in phonon dispersions (filled symbols)[37]. Modified from Figure 2 of reference[37].

3.4 Generality of approach

The approach described above for $MgB_2$ has common character with the topologies of other layered and non-layered superconducting compounds that display cosine-shaped electronic bands in close proximity to the Fermi level. For example, similar cosine shaped bands have been described for $CaC_6$, $LaH_{10}$ and β-$FeSe_{1-x}$.[7-10] For these binary compounds, we have shown the utility of DFT calculations as well as strategies, pertinent to the symmetry of each compound, to identify a superconducting gap.

TB equations for layered compounds are well known to result in cosine shaped bands.[6, 22, 28, 29] The results shown above and in other studies[4, 7-9] are strongly linked to the existence of cosine-shaped bands that show energy asymmetry that correlates with the known energy gap. To date, superconducting cuprates have not been analysed for asymmetry of cosine shaped bands, although they are discernible in publications of previous DFT calculations.[40, 41] Furthermore, the topology of the tubular shaped Fermi sections in many superconducting cuprates is favourable for nesting.

For superconducting solids of higher atomic number and greater element diversity, the required computational power will increase substantially. Accordingly, the resolution of DFT calculations used in this work may not suit accurate determination of asymmetry, for example, with cuprate superconductors. Nevertheless, theoretical hopping models that include cosine dependence, have provided valuable insights on superconductivity for the high $T_c$ cuprates.[24, 25] Compelling evidence for a hopping mechanism can be identified using DFT calculated, cosine shaped electronic bands that connect topological properties of these electronic bands and Fermi surfaces. While this discussion has focused on separate light and heavy effective mass σ bands and Fermi surfaces, the interplay between these bands *via* inter-band hopping is also likely to be present and relevant to superconducting mechanisms. This link between the geometry, or crystallography, of solids and their reciprocal space electronic properties at meV-scale is a key step towards predictable design of superconductors.

## 4. Conclusions

The EBS of materials provides a wealth of information on potential superconductivity mechanism(s) that cannot be extracted from a focus only on the density of states. For high symmetry structures (e.g. cubic, hexagonal, tetragonal), conversion to a lower symmetry (e.g. P1 derived from the atomic position space group) or consideration of superlattice(s) provides an EBS with lower band complexity compared to a higher symmetry calculation. For $MgB_2$, a 2*c* superlattice for EBS calculations is supported by spectroscopy experiments and previous DFT models.

*Ab initio* DFT calculations of $MgB_2$ show that the σ band along Γ–A is cosine-shaped; consistent with a modified TB model to account for energy asymmetry between principal nodes. Cosine band asymmetry is evident in calculated band structures for $MgB_2$ at ambient and higher pressures to 28.5 GPa. Bands calculated parallel to Γ–A and along the Γ–M direction provide details on the band structure at points within the Brillouin zone that approximate nodal inflection points for light and heavy effective mass σ bands. Bands along this direction, at or near $E_F$, split into two separate bands with an overlap at Γ of 29.0 meV. For bands calculated at further sampling intervals along Γ–M, the overlap at Γ reduces to 1.3 meV. Calculations at higher pressure show similar split bands near $E_F$ and with increased energy overlap at equivalent $k_y$ points. Electronic bands calculated for positions not at high symmetry nodes show that a hopping mechanism associated with coherent nesting occurs close to the Fermi surface.

Folded Fermi surfaces calculated from electronic bands for 2*c* superlattices of $MgB_2$ at different pressures provide criteria to identify disruption of nested coherent behaviour. The fraction of coherent nesting at calculated pressures, determined graphically using band intersections, enables estimates of $T_c$ that are consistent with experimental values for $MgB_2$. Estimated $T_c$ values determined by this alternative DFT-based method are also consistent with $T_c$ values obtained from measurement of the $E_{2g}$ phonon anomaly. The prevalence of cosine-shaped bands in superconductor EBSs as well as correlation between cosine band asymmetry and the superconducting gap suggest that use of a modified TB equation may be relevant to other compounds. To establish deterministic design criteria requires translation of reciprocal space EBSs to real space bonding-antibonding orbital configurations matched to DFT algorithms. The approach provided in this work may assist informed design of next generation superconducting compounds with targeted properties including room temperature transition.

## 5. Methods

DFT calculations on the EBS and FS for $MgB_2$ at pressures up to 30 GPa were undertaken with Materials Studio CASTEP Versions 2023 and 2025[42] following procedures detailed in earlier publications.[4, 11, 37] Key parameters such as plane wave cut-off energies, pseudopotentials, and Δ**k** grids are critical enablers of meV resolution for EBS and PD calculations. All band structure representations utilize .xlsx files converted from direct .csv outputs of Materials Studio DFT calculations. These data are then re-plotted using SigmaPlot for Windows Version 15 (Grafiti LLC, Palo Alto, CA USA) to enable precise determination of band energies and intersections at reciprocal lattice symmetry points.

Band intersections with high symmetry points as well as bands close to or at Fermi surfaces are identified by graphical interpretation and re-plotted as compiled figures. Band degeneracy, splitting and/or band avoidances are determined by visual inspection of SigmaPlot graphs correlated with energy values to the fifth significant figure (i.e. < 1meV). Precise interpretation of band intersections and/or avoidances can be obtained from DFT calculations undertaken with high resolution – a term used to describe the minimal input parameter values for DFT calculations as listed below.[4, 5, 11] Examples of high resolution calculations showing meV-scale evolution of bands along Γ–A with calculated changes to pressure are shown in **Figure S3** (Supporting Information).

All DFT calculations employ cut-off energies of 990 eV, Δ**k** grid of 0.005 $Å^{-1}$, norm-conserving pseudopotentials and either the Perdew-Zunger Local Density Approximation (LDA) or Generalised Gradient Approximation (GGA)[43, 44] for the exchange-correlation functional. Electronic minimisation parameters include a total energy/atom convergence tolerance of 0.5 x $10^{-6}$ eV, an eigen-energy convergence tolerance of 0.1375 x $10^{-6}$ eV, a Fermi energy convergence tolerance of 0.2721 x $10^{-13}$ eV and a Gaussian smearing scheme of width 0.1 eV. Optimisation calculations use the Broyden-Fletcher-Goldfarb-Shannon (BFGS) algorithm employing a total energy convergence tolerance of 0.5 x $10^{-5}$ eV. Fermi surfaces for the 1*c* primitive cell are used in order to reduce the clutter of folded Fermi surfaces that occur when calculated at higher pressures. Use of a primitive cell facilitates graphical extraction of Fermi surface projected coordinates and enables correct attribution to light or heavy effective mass Fermi surface(s).

All DFT outputs meet convergence criteria at the same fine tolerance level nominated for geometry optimization. The effects of input parameters to DFT calculations described in this work are not related to differences in calculation convergence, but critically, are due to the discreteness, or the finite number of reciprocal space points, used to select plane waves as basis functions. In earlier validation of our computational methods using *ab initio* DFT, parameters such as basis set cut-off values, Δk values, Fermi energy, an extensive array of functionals as well as software codes, have been varied and evaluated to optimise interpretation in reciprocal and real space.[5, 11, 45, 46] Optimised parameters are consistently used for a specific compound (e.g. $MgB_2$ and substituted $(Mg_{1-x}M_x)(B_{1-y}C_y)_2$).

We have utilised superlattices to confirm experimental spectroscopic data,[5, 20] to simplify computational methods for phonon dispersions of metal-substituted $MgB_2$,[38, 47] as well as to calculate a superconducting gap using EBSs of $MgB_2$.[4] Superlattice symmetry is maintained with a single primitive cell symmetry by the use of band folding in reciprocal space.[48, 49] The folded topologies of the electronic bands and Fermi surfaces, corresponding to the superlattice enable easier identification of electronic topological transitions and nesting relationships between extended sections of the Fermi surface.[4]

Acknowledgements

The authors acknowledge the e-Research Office at QUT for access to high performance computing. During the preparation of this manuscript and study, the authors did not use any artificial intelligence tools.

**Data Availability Statement**

Additional data presented in this study are available in Supporting Information. Raw data files from DFT calculations are available from the authors on request.

**References**

1. L. Pauling, The shared-electron chemical bond, *Proceedings of the National Academy of Sciences of the United States of America* **1928** 14 359–362.
2. L. Pauling, The Nature of the Chemical-Bond, *Journal of Chemical Education* **1992** 69 7 519–521.
3. J. M. An, W. E. Pickett, Superconductivity of $MgB_2$: Covalent bonds driven metallic, *Physical Review Letters* **2001** 86 19 4366–4369.
4. J. A. Alarco, I. D. R. Mackinnon, Superlattices, Bonding-Antibonding, Fermi Surface Nesting, and Superconductivity, *Cond Matter* **2023** 8 72 1–13.
5. J. A. Alarco, A. Chou, P. C. Talbot, I. D. R. Mackinnon, Phonon Modes of $MgB_2$: Superlattice Structures and Spectral Response, *Phys Chem Chem Phys* **2014** 16 24443–24456.
6. I. N. Askerzade, Reviews of Topical Problems - Study of layered superconductors in the theory of an electron - phonon coupling mechanism, *Physics - Uspekhi* **2009** 52 10 977–988.
7. J. Alarco, I. Mackinnon, Spin Orbit Coupling for Superconductivity Models of $LaH_{10}$, *Annalen Der Physik* **2026** 538 1–17.
8. B. C. Wang, A. Bianconi, I. D. R. Mackinnon, J. A. Alarco, Superlattice Symmetries Reveal Electronic Topological Transition in $CaC_6$ with Pressure, *Crystals* **2024** 14 554 1–21.
9. B. C. Wang, A. Bianconi, I. D. R. Mackinnon, J. A. Alarco, Superlattice Delineated Fermi Surface Nesting and Electron-Phonon Coupling in $CaC_6$, *Crystals* **2024** 14 499 1–17.
10. I. D. R. Mackinnon, J. A. Alarco, Cosine bands, flat bands and superconductivity in orthorhombic iron selenide, *Annalen der Physik* **2026** in review.
11. I. D. R. Mackinnon, Almutairi A., J. A. Alarco, Insights from systematic DFT calculations on superconductors, in: Arcos, J.M.V. (Ed.), *Real Perspectives of Fourier Transforms and Current Developments in Superconductivity*, IntechOpen Ltd., London UK, **2021**, pp. 1–29.
12. R. Heid, K. P. Bohnen, B. Renker, *Electron-phonon coupling and superconductivity in $MgB_2$ and related diborides*, Spring Meeting of the Deutsche-Physikalische-GesellschaftRegensburg, Germany, 2002, pp. 293–305.

13. K. Kunc, I. Loa, K. Syassen, R. Kremer, K. Ahn, $MgB_2$ under pressure: phonon calculations, Raman spectroscopy, and optical reflectance*, J. Phys.: Condens. Matter* **2001** 13 44 9945–9962.
14. A. Q. R. Baron, H. Uchiyama, S. Tsutsui, et al., Review: Phonon spectra in pure and carbon doped $MgB_2$ by inelastic X-ray scattering*, Physica C* **2007** 456 83–91.
15. J. Kortus, Current progress in the theoretical understanding of $MgB_2$*, Physica C: Superconductivity* **2007** 456 54–62.
16. J. A. Alarco, P. C. Talbot, I. D. R. Mackinnon, Electron density response to phonon dynamics in $MgB_2$: an indicator of superconducting properties*, Mod. Numer. Sim. Mater. Sci.* **2018** 8 21–46.
17. E. Canadell, M.-L. Doublet, C. Lung, *Orbital Approach to the Electronic Structure of Solids*, Oxford University Press, Oxford, UK, **2012**.
18. E. Canadell, M. Whangbo, Conceptual aspects of structure property correlations and electronic instabilities, with applications to low-dimensional transition-metal oxides*, Chemical Reviews* **1991** 91 5 965–1034.
19. T. Albright, J. Burdett, M. Whangbo, *Orbital Interactions in Chemistry*, 2nd Edition ed., John Wiley & Sons, Inc.**2013**.
20. J. A. Alarco, B. Gupta, M. Shahbazi, D. Appadoo, I. D. R. Mackinnon, THz/Far infrared synchrotron observations of superlattice frequencies in $MgB_2$*, Phys Chem Chem Phys* **2021** 23 41 23922–23932.
21. C. Buzea, T. Yamashita, TOPICAL REVIEW- Review of the superconducting properties of $MgB_2$*, Supercond. Sci. Technol.* **2001** 14 R115–R146.
22. I. Askerzade, *Unconventional Superconductors - Anisotropy and Multiband Effects*, Springer-Verlag, Berlin Heidelberg, **2012**.
23. C. Kittel, *Introduction to solid state physics*, 8th ed., John Wiley & Sons, Inc, United States of America, **2005**.
24. A. Mourachkine, *High-Temperature Superconductivity in Cuprates*, Kluwer Academic Publishers, Dordrecht, **2002**.
25. S.-i. Uchida, *High Temperature Superconductivity*, Springer, Tokyo Japan, **2015**.
26. W. Kohn, Nobel Lecture: Electronic structure of matter—wave functions and density functionals*, Rev. Modern Phys.* **1999** 71 5 1253–1266.
27. Y. Kong, O. V. Dolgov, O. Jepsen, O. K. Andersen, Electron-phonon interaction in the normal and superconducting states of $MgB_2$*, Phys. Rev. B* **2001** 64 020501 1–5.
28. N. W. Ashcroft, N. D. Mermin, *Solid State Physics*, Saunders, Philadelphia PA USA, **1976**.
29. P. Hofmann, *Solid State Physics: An Introduction*, Third ed., Wiley VCH, Berlin Germany, **2022**.
30. M. Whangbo, S. Deng, J. Köhler, A. Simon, Interband Electron Pairing for Superconductivity from the Breakdown of the Born-Oppenheimer Approximation*, Chemphyschem* **2018** 19 23 3191–3195.
31. M. Whangbo, On the Relative Stability of the Metallic and the Insulating States of a Half-filled Band*, Journal of Chemical Physics* **1980** 73 8 3854–3861.
32. M. D. Johannes, I. I. Mazin, Fermi surface nesting and the origin of charge density waves in metals*, Phys. Rev. B* **2008** 77 16 165135.
33. W. E. Pickett, Colloquium: Room temperature superconductivity: The roles of theory and materials design*, Reviews of Modern Physics* **2023** 95 2 1–29.
34. A. P. Sutton, *Electronic Structure of Materials*, Clarendon Press: Oxford Science Publications Oxford UK **2004**.
35. J. Tang, L.-C. Qin, A. Matsushita, et al., Lattice parameter and $T_c$ dependence of sintered $MgB_2$ superconductor on hydrostatic pressure*, Phys. Rev. B* **2001** 64 132509 1–4.
36. S. Deemyad, T. Tomita, J. Hamlin, et al., Dependence of the superconducting transition temperature of single and polycrystalline $MgB_2$ on hydrostatic pressure*, Physica C: Superconductivity* **2003** 385 1 105–116.

37. J. A. Alarco, P. C. Talbot, I. D. R. Mackinnon, Phonon dispersion models for $MgB_2$ with application of pressure, *Physica C: Supercond Appl* **2017** 536 11–17.
38. J. A. Alarco, P. C. Talbot, I. D. R. Mackinnon, Phonon anomalies predict superconducting $T_c$ for $AlB_2$-type structures, *Phys Chem Chem Phys* **2015** 17 38 25090–25099.
39. P. Malhotra, *Analytical Chemistry: Basic Techniques and Methods*, Springer Nature Switzerland AG, Cham, Switzerland, **2023**.
40. D. J. Singh, Electronic-structure of $HgBa_2CuO_4$, *Physica C* **1993** 212 1-2 228–232.
41. K. H. Bennemann, J. B. Ketterson, *The Physics of Superconductors: Volume II*, Springer Berlin, Heidelberg, Germany, **2004**.
42. S. J. Clark, M. D. Segall, C. J. Pickard, et al., First principles methods using CASTEP, *Z Kristallogr* **2005** 220 5-6 567–570.
43. J. P. Perdew, K. Burke, M. Ernzerhof, Generalized Gradient Approximation Made Simple, *Phys Rev Lett* **1996** 77 18 3865–3868.
44. J. P. Perdew, J. A. Chevary, S. H. Vosko, et al., Atoms, molecules, solids, and surfaces: Applications of the generalized gradient approximation for exchange and correlation, *Phys Rev B* **1992** 46 11 6671–6687.
45. J. A. Alarco, I. D. R. Mackinnon, Phonon dispersions as indicators of dynamic symmetry reduction in superconductors, in: Stavrou, V.N. (Ed.), *Phonons in Low Dimensional Structures*, InTech Open, London UK, **2018**, pp. 75–101.
46. J. A. Alarco, P. C. Talbot, I. D. R. Mackinnon, Comparison of functionals for metal hexaboride band structure calculations, *Mod. Numer. Sim. Mater. Sci.* **2014** 4 53–69.
47. I. D. R. Mackinnon, P. C. Talbot, J. A. Alarco, Phonon dispersion anomalies and superconductivity in metal substituted $MgB_2$, *Comp. Mater. Sci.* **2017** 130 191–203.
48. H. Jones, *The Theory of Brillouin Zones and Electronic States on Crystals*, North-Holland Publishing Company Amsterdam, **1960**.
49. S. L. Altmann, *Band Theory of Solids: An Introduction from the Point of View of Symmetry*, Clarendon Press, Oxford **2002**.

**Supporting Information**

Supporting Information is available from the Wiley Online Library or from the authors.